# Limiting Efficiencies of Multiple Exciton Generation Solar Cells with Quantum Yield Extraction

**Jongwon Lee**

Independent Researcher, Seoul, Republic of Korea
E-mail: jlee159@asu.edu

**Abstract:**[1] Multiple exciton generation solar cells exhibit a low power conversion efficiency owing to nonradiative recombination even if numerous electron and hole pairs are generated per incident photon. This paper elucidates the non-idealities of multiple exciton generation solar cells (MEGSCs). First, we present mathematical approaches for determining the quantum yield (QY) to discuss the non-idealities of MEGSCs by adjusting the delta function. In particular, we employ the Gaussian distribution function to present the occupancy status of carriers at each energy state by Dirac delta function. By adjusting the Gaussian distribution function for each energy state, we obtain the ideal and non-ideal QYs. Through this approach, we discuss the material imperfections of MEGSCs by analyzing the mathematically obtained QYs. By calculating the ratio between the radiative and nonradiative recombination, we can discuss the status of radiative recombination. Furthermore, we apply this approach into the detailed balance limit of MEGSC to investigate the practical limit of MEGSC.

**Keywords:** detailed balance limit; multiple exciton generation solar cell; quantum yield; delta function

## 1. Introduction

Multiple exciton generation solar cells (MEGSCs) are promising photovoltaic devices for surpassing the Shockley–Queisser limit [1]. This is because they entail the creation of multiple electron and hole pairs (EHPs). One incident photon with an energy (=$E_{ph}$) greater than the bandgap energy (=$E_g$) can generate multiple EHPs owing to impact ionization in bulk or nanostructured solar cells [2,3]. The number of generated and recombined electrons should be conserved under substantial blackbody radiation. Since Kolondinski et al.[2] first reported the MEG effect (quantum efficiency of over 100%) in bulk Si solar cells, advanced theoretical research has suggested the use of nanostructures (e.g., quantum dots) because they offer excellent quantum confinement [3,4]. Furthermore, there have been various studies on the efficiency limit of MEG and the possibility of its occurrence in nanostructures. In such studies, a quantum yield (QY) of over 100% has been observed once the threshold energy (=$E_{th}$) was exceeded [5-11]. Notably, QY and $E_{th}$ are critical parameters that determine the ideality of MEGSCs [10,11]. Here, $E_{th}$ refers to the photon energy at which the QY exceeds 100%. In theory, QY increases in accordance with a staircase (step) function. However, the experimental data regarding QY deviate from the ideal case: QY increases linearly after $E_{th}$ [10, 11]. In other words, QY depends on the material quality (surface states) and the loss of carriers corresponding to each energy state in quantum dot structures [12-16]. The non-ideal QY (NQY) is experimentally extracted via pump-probe measurements by comparing the peaks corresponding to the excited and ground states [10-16]. Further, it is theoretically modeled using the MEG efficiency (=$E_g$/The additional energy required to create further EHPs), which in turn can be employed for calculating the slope of the NQY. In addition, for NQY, $E_{th}$ is also defined by the effective mass and MEG efficiency [17]. These results have prompted further studies, such as those regarding (1) the theoretical efficiency with the corresponding NQY and $E_{th}$ and (2) the extraction of the NQY via pump-probe measurements and statistical approaches [10-14,18-20,21-23]. In mathematics, the derivative of the Heaviside step function is the Dirac

---

[1] EHP: electron and hole pairs;  IQY; ideal quantum yield; MEGSC: multiple exciton generation solar cells; NQY; non-ideal quantum yield; QY: quantum yield

delta function [24]. Thus, the derivative of the ideal QY (IQY) may yield the Dirac delta function at $m \cdot E_g$, where m is a positive integer. And, the integration of the Dirac delta function yields the step function. Thus, it could be possible to determine the generalized QY based on the variation of the delta function.

In this regard, this work employs mathematics to explain the material dependence of the performance/characteristics of MEGSCs. The development of near-perfect nanostructured materials is crucial because there exist non-idealities in MEGSCs [25]. Notably, one major limitation of MEGSCs is that their power conversion efficiency is substantially lower than the expected theoretical efficiency owing to material imperfections, although substantial research has been conducted in this regard [25-28]. Therefore, further development of nanostructured solar cells with enhanced material quality is necessary to ensure that negligible defects occur in MEGSCs.

First, this paper discusses the Dirac delta function and the unit step function; the IQY can be explained based on their mathematical relationship: the integration of Heaviside unit step functions under a shift of $m \cdot E_g$ results in IQY and their derivative yields a group of delta-functions. Further, the deviation exhibited by the delta function can describe the NQY. Most QY extractions are performed by fitting certain parameters based on experimental results [10-16]. In contrast, this study employs a purely theoretical approach by adjusting the delta function to obtain both IQY and NQY; moreover, it analyzes the non-idealities of MEGSCs. These aspects have not been sufficiently addressed in prior research. Although this approach is mathematical, this simple yet novel method has not been utilized to explain the material quality in MEGSCs. The NQY is a good indicator of material status in quantum dot MEGSCs as it includes material information. Notably, even if substantial experimental achievements are made regarding the QY, it will remain difficult to implement efficient MEGSCs without a near-perfect solar cell material.

The conventional detailed balance (DB) limit considers only one-radiative recombination to predict the maximum theoretical efficiency limit. However, NQY includes the materials information such as nonradiative recombination (NR) (Auger recombination, surface defects and phonon) due to loss of carriers. Therefore, the DB limit of MEGSC has to include it. Thus, we use this proposed QY extraction method into the DB limit of MEGSC to predict the more practical efficiency limit by comparing peak-intensities (maximum peak at $E_g$ and $2^{nd}$ maximum peak at $2E_g$) while considering NR impact. In other words, the ratio between peak intensities describes the status of radiative recombination while including NR and its correlated an ideal reverse saturation current density. Therefore, the variation of ideal saturation current can determine the performance of MEGSC.

In this paper, we present a theoretical approach for the determination of QY based on the delta function for each energy state in nanostructured MEGSCs. Next, we discuss the approaches for extracting QY and its non-idealities. Finally, we apply these concepts into the DB limit of MEGSC with NR to discuss the practical limit of MEGSC.

## 2. Theory

In theory, the derivative of the unit step function is the Dirac delta function. Thus, the staircase function can be written as follows [29]

$$IQY = H(E-E_g) + H(E-2 \cdot E_g) + H(E-3 \cdot E_g) + \cdots + H(E-m \cdot E_g) + \cdots + H(E-M \cdot E_g) \quad (1)$$

where H is the Heaviside step function, m is a positive integer, and $E_g$ is the bandgap energy

Furthermore, the derivative of each step function is

$$\delta = [\delta(E-E_g), \delta(E-2 \cdot E_g), \delta(E-3 \cdot E_g), \ldots, \delta(E-m \cdot E_g), \ldots, \delta(E-m \cdot E_g)] \quad (2)$$

where δ is the delta function

The Dirac delta functions can be employed to represent the discrete energy states in quantum dots. For this simulation, the author assumes that if multiple electrons corresponding to each energy state $m \cdot E_g$ certainly exist, the Dirac delta function at $m \cdot E_g$ can be represented the occupancy status of electrons. Consequently, this means that the material structure for MEGSCs can exhibit excellent idea qualities. To mathematically describe delta functions, we use a Gaussian distribution function (GDF), as shown in Eq. (3). Furthermore, we set the mean or expected value of the variable corresponding to the distribution shown in Eq. (3) as $m \cdot E_g$ to represent the multiple energy states.

$$f(E) = \frac{1}{\sigma\sqrt{2\pi}} e^{-\frac{1}{2}(\frac{E-m\cdot E_g}{\sigma})^2} \qquad (3)$$

where σ is the standard deviation, (=0.004, 0.1, 0.2, 0.3, 0.4, 0.5…), E is an arbitrary energy, $E_g$ is the bandgap energy, and m=1,2,3,…

First, we calculate GDF for $m\cdot E_g$, where m=1,2,3⋯ (Eq.(3), Eq.(4) and Eq.(5)), to obtain the Dirac delta function corresponding to each energy state. Note that $f_1(E)$ is the GDF at $E_g$, which corresponds to the generation of one EHP. Next, we employ cumulative integration and normalize the value of E (Eq.(5)) at each energy state. Finally, we add the values to obtain QY (Eq.(7)). This procedure is illustrated in Fig. 1 for IQY and in Fig. 2 and Fig. 3 for NQY.

$$f_1(E) = \frac{1}{\sigma_1\sqrt{2\pi}} e^{-\frac{1}{2}\left(\frac{E-E_g}{\sigma_1}\right)^2} \qquad (4)$$

$$f_m(E) = [f_1(E), \frac{1}{\sigma_2\sqrt{2\pi}} e^{-\frac{1}{2}\left(\frac{E-2\cdot E_g}{\sigma_2\sigma}\right)^2}, \frac{1}{\sigma_3\sqrt{2\pi}} e^{-\frac{1}{2}\left(\frac{E-3\cdot E_g}{\sigma_3}\right)^2}, \cdots]/f_1(E) \qquad (5)$$

$$f_n(E) = \begin{bmatrix} \frac{\sum_{i=1}^{n}\int_{E_i=0\,eV}^{x_n} f_1(E)\,dE}{\int_{E_i=0\,eV}^{E_f=10\,eV} f_1(E)dE}, \frac{\sum_{i=1}^{n}\int_{E_i=0\,eV}^{x_n} \frac{1}{\sigma_2\sqrt{2\pi}} e^{-\frac{1}{2}\left(\frac{E-2\cdot E_g}{\sigma_2}\right)^2} dE}{\int_{E_i=0\,eV}^{E_f=10\,eV} f_1(E)dE}, \cdots \\ \frac{\sum_{i=1}^{n}\int_{E_i=0\,eV}^{x_n} \frac{1}{\sigma_3\sqrt{2\pi}} e^{-\frac{1}{2}\left(\frac{E-3\cdot E_g}{\sigma_3}\right)^2} dE}{\int_{E_i=0\,eV}^{E_f=10\,eV} f_1(E)dE}, \cdots \end{bmatrix} \qquad (6)$$

$$\text{IQY} = \text{sum}(f_m(E)) \qquad (7)$$

where $f_1$ is the GDF at $E_g$, $f_m(E)$ is that at $m\cdot E_g$, $-\infty < x_n < \infty$, σ is the standard deviation corresponding to the energy state (=$m\cdot E_g$), m=1,2,3⋯, and $f_n$ represents the calculation of the unit-step function from $f_m$ (Eq (3)) at each energy state. Moreover, $E_i$ is the initial energy state, $E_f$ is the final energy state, and $x_n$ is 0.01, 0.02,…10 eV.

As σ decreases, the GDF will exhibit a highly sharp peak, similar to a delta function. In contrast, increasing σ increases the width and lowers the peak of the GDF distribution. Therefore, the deviation from the ideal condition (i.e., the non-ideal condition) can be effectively described by varying σ; consequently, the non-idealities of materials can be explained. Thus, here we employ a small value of σ and show the high peak at $E_g$ to explain the strong evidence of the generation of one EHP. Next, the integration of each delta function yields the unit-step function at each energy state. Finally, the cumulative sum of each step function produces the IQY.

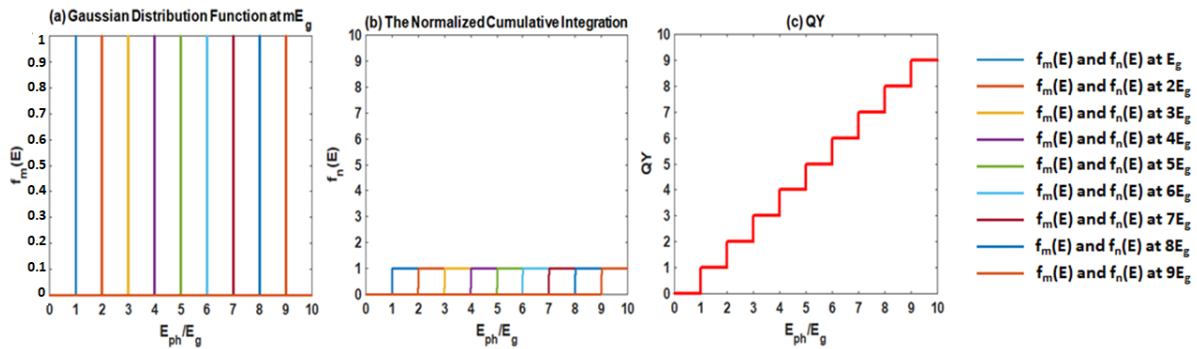

Fig. 1 Conversion from the Gaussian distribution function to the IQY, (a) represents the delta function by normalized GDF of each energy state, (b) is the step function of each energy state, and (c) shows the corresponding QYs. The units are arbitrary for (a), (b), and (c).

To determine the value of σ for each energy state, we refer to the value of external radiative recombination efficiency (the ratio between radiative and non-radiative recombination) or the fraction of radiative recombination, $f_C$, to determine the peak-intensity corresponding to each energy state [1,30,31]. A small external radiative efficiency (ERE) or $f_C$, such as $10^{-5}$–$10^{-10}$, indicates a high non-radiative recombination rate, which in turn leads to a low power conversion efficiency [30]. Thus, we use this value to determine the ratio between the second and first peaks to represent the high non-radiative recombination rate. For instance, with regard to IQY, we assume σ as $4 \times 10^{-3}$ to obtain a normalized sharp peak (=1) at each energy state, similar to the delta function in a zero-dimensional structure (see Fig. 1 (a) and Eq. (5)). The normalized cumulative integration of this function yields an individual step function for each state (see Fig. 1 (b) and Eq. (6)), and the summation of each step function yields the staircase function (see Fig. 1 (c) and Eq. (7)). In contrast, NQY entails an increased $E_{th}$ and a different shape of the QY curve. The increased $E_{th}$ in turn necessitates increased photon energy to generate multiple EHPs. In addition, NQY involves the loss of carriers corresponding to each energy state owing to non-radiative recombination. To mathematically describe NQY, we initially assume that the generation of one EHP occurs at $E_g$, corresponding to the sharp peak ($\sigma_1=4\times10^{-3}$) of the GDF. Next, we choose $\sigma_m=0.1$ at $m \cdot E_g$ ($m=2,3,4\cdots$) to account for the low occupancy corresponding to electron confinement for each energy state (Fig. 2 (a) and Eq. (8)) based on ERE or $f_C$. As $\sigma_m$ decreases, the delta function loses its characteristics in that the peak widens and decreases. In this study, the author set the value $\sigma_2=0.1$ based on the investigation of ERE and $f_C$ [1,30,31]. In this case, the ratio "peak intensity at $2E_g$/peak intensity at $E_g$" is approximately 0.04, which represents the high non-radiative recombination rate of MEGSCs. To calculate the increased $E_{th}$, we choose a point corresponding to $f(E) \geq 0$ at $2E_g$, which lies on the left-side of the GDF distribution ($=E_d=2E_g-x$) (see Fig. 2 (b)). We append this point at $m \cdot E_g$; therefore, the peaks are redistributed at $m \cdot E_g+E_d$. Thus, adjusting σ shows the deviation from ideality, and its shifted peak shows points adjusted for the occupancy of carriers. Cumulative integration with respect to the normalized values of E in this distribution for various energy states (see Fig. 2 (c) and Eq. (9)) and the summation of the resultant values shows that $E_{th} \approx 2.4E_g$ for NQY (see Fig. 2 (d) and Eq. (10)). In Fig. 2 (c), the step function corresponding to $2E_g+E_d$ exhibits a deviation, which represents the loss of carriers. In Fig. 2, after $E_{th}$, the step function starts to lose its characteristics owing to the loss of carriers.

$$f'_m(E) = [f_1(E), \frac{1}{\sigma_2\sqrt{2\pi}} e^{-\frac{1}{2}\left(\frac{E-(2 \cdot E_g+E_d)}{\sigma_2}\right)^2}, \cdots, \frac{1}{\sigma\sqrt{2\pi}} e^{-\frac{1}{2}\left(\frac{E-(m \cdot E_g+E_d)}{\sigma_3}\right)^2} \cdots] / f_1(E) \tag{8}$$

$$\text{normalized } f'_m(E) = \begin{bmatrix} \dfrac{\sum_{i=1}^n \int_{E_i=0\ eV}^{x_n} f_1(E)\, dE}{\int_{E_i=0\ eV}^{E_f=10\ eV} f_1(E)dE}, \dfrac{\sum_{i=1}^n \int_{E_i=0\ eV}^{x_n} \frac{1}{\sigma_2\sigma\sqrt{2\pi}} e^{-\frac{1}{2}\left(\frac{E-(2 \cdot E_g+E_d)}{\sigma_2}\right)^2} dE}{\int_{E_i=0\ eV}^{E_f=10\ eV} f_1(E)dE}, \cdots \\ \dfrac{\sum_{i=1}^n \int_{E_i=0\ eV}^{x_n} \frac{1}{\sigma_3\sqrt{2\pi}} e^{-\frac{1}{2}\left(\frac{E-(m \cdot E_g+E_d)}{\sigma_3}\right)^2} dE}{\int_{E_i=0\ eV}^{E_f=10\ eV} f_1(E)dE}, \cdots \end{bmatrix} \tag{9}$$

$$\text{NQY} = \text{sum (normalized } f_m'(E)) \tag{10}$$

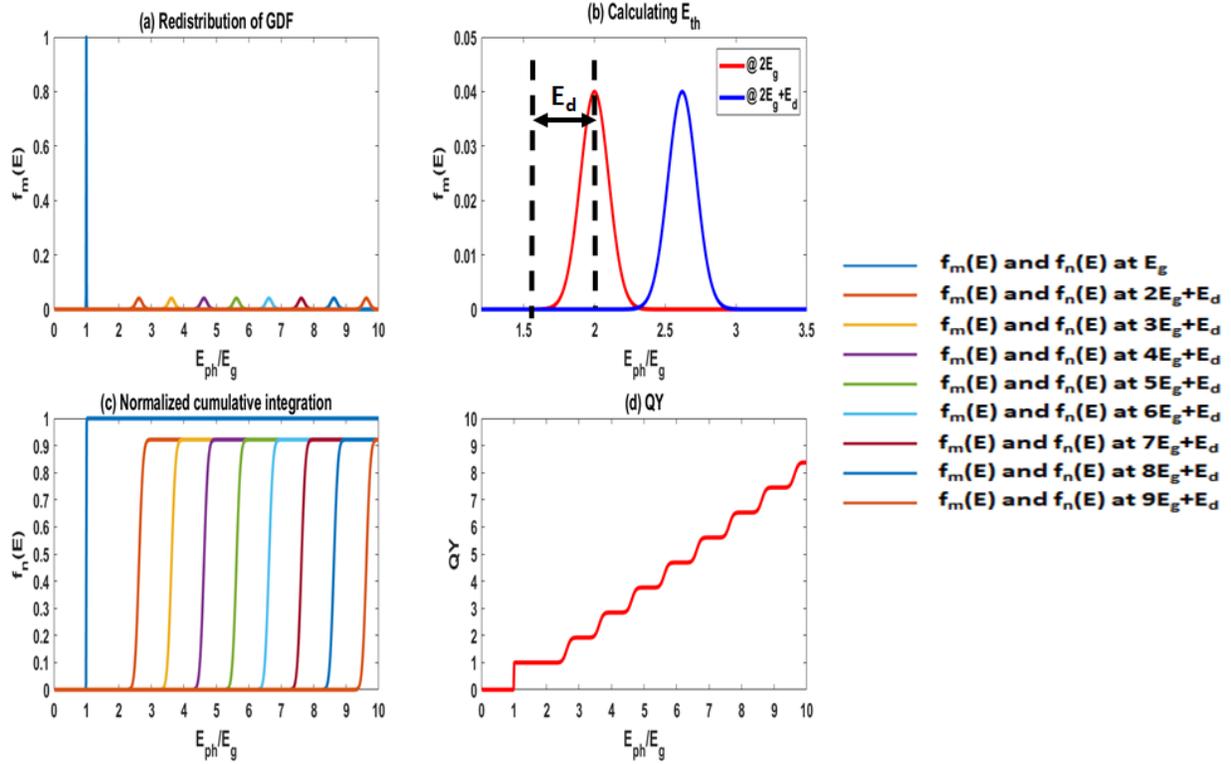

Fig.2 Conversion from the Gaussian distribution function to the NQY at σ₁=0.004 at $E_g$ and $σ_m$=0.1 at m·$E_g$ (m=2,3,4,…). Here, (a) displays the normalized and redistributed GDF, (b) shows the corresponding GDF distribution for calculating $E_{th}$, (c) illustrate the results of Eq (7) to represent the deviation from the ideal case, and (d) shows the corresponding QY. The units are normalized for (a) and (b).

And, the author reorganize the DB of MEGSC by adjusting reverse saturation current density ($J_0$) by including non-non-radiative recombination ratio. For this approach, we define $f_{NR}$ which it is the ratio of between the maximum value of $f_1(E_g)$ and $f_2'(2E_g+E_d)$. And, $f_{NR}$ represents a measure of radiative recombination while including NR term. And, its equations are shown in Eqns. (11)-(16)

$$f_{NR} = \frac{\max(f_2(2E_g + E_d))}{\max(f_1(E_g))} = \frac{\max\left(\frac{1}{\sigma_2\sqrt{2\pi}} e^{-\frac{1}{2}\left(\frac{E-(2\cdot E_g+E_d)}{\sigma_2}\right)^2}\right)}{\max\left(\frac{1}{\sigma_1\sqrt{2\pi}} e^{-\frac{1}{2}\left(\frac{E-(E_g)}{\sigma_1}\right)^2}\right)} \quad (11)$$

$$\phi_{MEG}(E_1, E_2, T, \mu) = \frac{2\pi}{h^3 c^2} \int_{E_1}^{E_2} \frac{QY(E) \cdot E^2}{\exp[(E-\mu_{MEG})/kT]-1} dE \quad (12)$$

$$J_{BB} = q \cdot C \cdot f_s \cdot \phi_{MEG}(E_g, \infty, T_S, 0) \\ + q \cdot C \cdot (1-f_s) \cdot \phi_{MEG}(E_g, \infty, T_C, 0) \\ - q \cdot \phi_{MEG}(E_g, \infty, T_C, \mu_{MEG}) \quad (13)$$

$$\mu_{MEG} = q \cdot QY(E) \cdot V \quad (14)$$

where ϕ is the particle flux given by Planck's equation for a temperature T with a CP μ in the photon energy range between $E_1$ and $E_2$. h is Planck's constant, c is the speed of light, and μ is the CP of an SJSC (q·V), where V is the operating voltage. $μ_{MEG}$ is the CP of MEG (q·QY(E)·V), k is the Boltzmann constant, J is the current density of the solar cell, q is the element of the charge, C is the optical concentration, $f_S$ is a geometry factor (1/46200), $T_S$ is the temperature of the sun (6000K), and $T_C$ is the solar cell's temperature (300K).

From DB limit theory [1,31], the nonradiative and radiative generation and recombination should be conserved by controlling non-radiative recombination ratio. And, its balanced equation is Eq. (15) and an organized form is shown in Eq. (16) [1,31]

$$F_{S,MEG} - F_{C,MEG}(V) + R_{MEG}(0) - R_{MEG}(V) - J_{BB}/q = 0 \tag{15}$$

where $F_{S,MEG}$ and $F_{C,MEG}(V)$ are, respectively, the generation and recombination for the radiative term, and $R_{MEG}(0)$ and $R_{MEG}(V)$ are the non-radiative generation and recombination, respectively [31].

$$\begin{aligned} J_{BB} &= q \cdot (F_{S,MEG} - F_{C0,MEG}) \\ &+ q \cdot (F_{C0,MEG}/f_{NR}) \left[ 1 - \exp\left( \frac{q \cdot QY \cdot V}{k \cdot T_C} \right) \right] \\ &= q \cdot (F_{S,MEG} - F_{C0,MEG}) + J_0 \cdot \left[ 1 - \exp\left( \frac{q \cdot QY \cdot V}{k \cdot T_C} \right) \right] \end{aligned} \tag{16}$$

where $F_{S,MEG} = C \cdot f_S \cdot \phi_{MEG}(E_g,\infty,T_S,0) + C \cdot (1-f_S) \cdot \phi_{MEG}(E_g,\infty,T_C,0)$, $F_{C0,MEG} = \phi_{MEG}(E_g,\infty,T_C,0)$

$f_{NR}=1$ describes the ideal condition of MEGSC due to only considering radiative recombination. And, $f_{NR}$ is inversely proportional to $J_0$. For instance, $f_{NR}=0.04$ means that the ideal saturation current is increased by 25 times of $J_0$ due to including NR term and its theoretical efficiencies will be decreased. And, we will discuss this in the results section.

## 3. Results

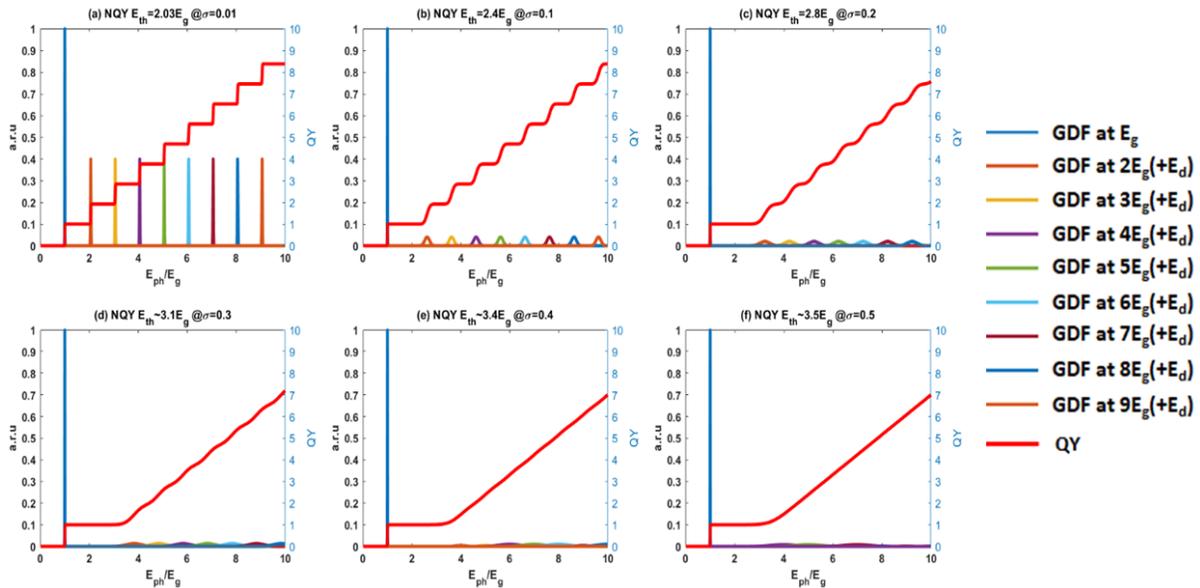

Fig. 3 QY extraction from GDF distribution with variation of σ. From (a) to (f), we normalized the value of GDF to present the probability of carrier occupancy. As increasing σ, the peak intensity and width of GDF are lowered and extended which it present the low occupancy rate of carriers at each energy state and its threshold energy is increased

In this section, we discuss the variation of σ and the non-idealities of QY and its corresponding DB limit of MEGSC with including NR impact. In ideal nanostructured MEGSCs, QY is given by the staircase function because there are no carrier losses owing to the near-ideal material quality. However, most measurement data show that QY exhibits a linear increase once $E_{th}$ becomes greater than $2E_g$ owing to material defects (e.g., surface traps) [12-16]. Fig. 3 presents the results of the calculation of QY based on the GDF distribution (the occupancy of carriers). When σ = 4 × $10^{-3}$ for every energy state, QY exhibits a staircase curve, similar to that exhibited by IQY (see Fig. 1). However, as σ increases, the shape of the delta function plot begins losing its characteristics due to loss of carriers at zero-dimensional structures. That is, the probability of occupation of energy states by carriers becomes low for every energy state. Therefore, we could regard this as the effect of delayed energy states. Finally, the shape of the QY plot becomes linear with increased $E_{th}$. (see Fig. 3 (a)-(f)). For instance, in Fig. 3, the QY shapes for σ=0.01, 0.1 and 0.2 (Fig, 3(a)-(c)) do not accurately correspond to a step function, and $E_{th}$ is below $3E_g$. When σ=0.3 (Fig. 3 (d)), $E_{th}$ is found to be equal to $3.1E_g$, and the staircase function plot does not maintain its characteristic shape. Typically, when σ is 0.01 (see Fig.3(a), staircase step function still shows but its maximum value at 10 is less than 9 of IQY case. Furthermore, when σ=0.4 and 0.5 (See Fig. 3 (e) and (f)), QY increases linearly, exhibiting slightly less than 1 (=slope), and $E_{th}$ = 3.4~$3.5E_g$. This indicates that MEGSCs start to behave similar to conventional single junction solar cells owing to the increased $E_{th}$. Thus, the width and height of the peak of the GDF distribution for each energy state indicate the status of non-ideality that the characteristics of the delta function describe, which in turn is related to the material imperfections of nanostructures. In other words, the IQY case entails high possibility of carriers occupying energy levels or maintaining a sharp peak for each energy state (=m·$E_g$). In addition, the material quality of the nanostructure is well-maintained, without carrier loss. However, NQY describes the imperfections of MEGSCs due to NR such as those arising from the occurrence of surface traps inside nanostructures [12-16]. The peak intensity exhibited by the GDF is a key parameter in determining the quality of nanostructures. Except for the IQY case, the peaks (from $2E_g$) indicate low occupancies of carriers. For instance, when the highest peak at $E_g$ corresponds to 1, the second highest peak is below 0.04 (at σ=0.1). The intensity of the highest peak at $E_g$ is approximately 25 times higher than that of the second highest peak at $2E_g$, indicating that the carrier generation rate at m·$E_g$ is not sufficiently high owing to NR or material defects. This difference gives rise to a substantial reverse saturation current because of the increasing non-radiative recombination rate, thereby degrading electrical parameters such as the open circuit voltage and carrier losses. Thus, the curve exhibited by QY starts to lose its staircase shape; furthermore, it becomes linear owing to carrier losses at each energy state. Therefore, the NR ratio is increased. For the successful demonstration of the advantages of nano-structure in MEGSCs, the solar cell material should be nearly perfect, with negligible non-radiative recombination, to surpass the SQ limit.

|  | without $f_{NR}$ | | with $f_{NR}$ | | |
|---|---|---|---|---|---|
|  | $E_g$ (eV) | Efficiency (%) | $f_{NR}$ | $E_g$ (eV) | Efficiency (%) |
| IQY (σ=0.004) | 0.77 | 44.6 | N/A | N/A | N/A |
| σ=0.1, $E_{th}$ = 2.03 $E_g$ | 1.01 | 41.5 | 0.40 | 1.01 | 40.1 |
| σ=0.1, $E_{th}$ = 2.4 $E_g$ | 1.03 | 34.5 | 0.040 | 1.14 | 30.8 |
| σ=0.2, $E_{th}$ = 2.8 $E_g$ | 1.20 | 31.8 | 0.020 | 1.31 | 28.3 |
| σ=0.3, $E_{th}$ = 3.1 $E_g$ | 1.28 | 31.2 | 0.0133 | 1.37 | 27.6 |
| σ=0.4, $E_{th}$ = 3.4 $E_g$ | 1.30 | 31.0 | 0.01 | 1.39 | 27.3 |
| σ=0.5, $E_{th}$ = 3.5 $E_g$ | 1.31 | 31.0 | 0.008 | 1.40 | 27.2 |

Table.1 optimum bandgaps and its maximum efficiencies with/without NR under one sun illumination

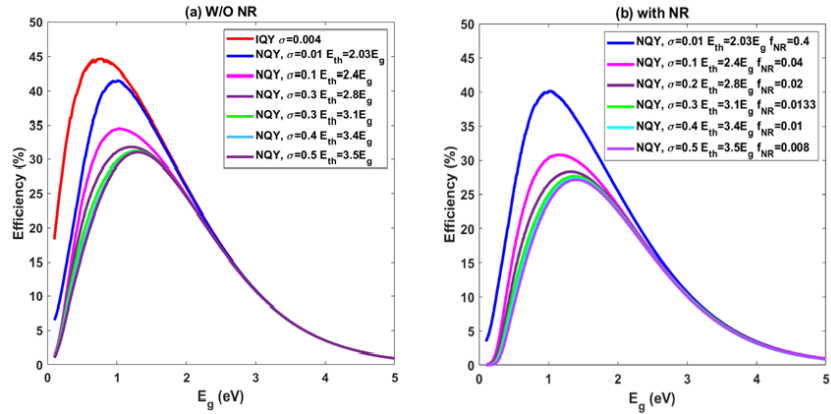

Fig. 4 Efficiency vs $E_g$ after DB calculation with/without NR recombination impact under one sun illumination. The theoretical efficiencies are saturated after σ=0.4

To compare different slope of NQY, we also provide different $σ_1=2×10^{-3}$ (see Fig.5 (a) and (b), Table 2). For this simulation, after increasing two times of σ, its corresponding slope at σ=0.5 is 0.49. This value is approximately 50% lower than the case of $σ=4×10^{-3}$. Except $σ_1$=0.01, other theoretical efficiencies are lowered than 30% which it could not surpass the SQ limit (see Fig.5 (b)). In other words, even if there are QY over 100% after $E_{th}=2E_g$, the MEGSC become weak or lost its performances due to its NR impact by increased reverse saturation current densities. Only $E_{th}=2.03E_g$ case provides the expected results of MEGSC that $E_{th}$ should be nearly closed to $2E_g$ and its QY shape has to be staircase step function even if its value is not ideal case.

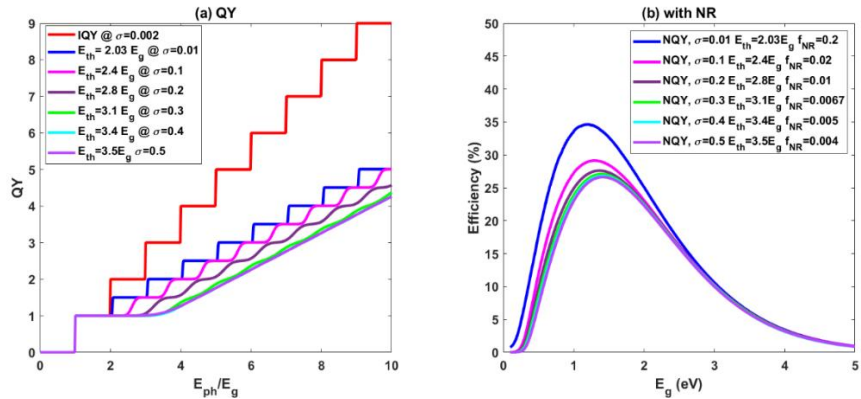

Fig. 5 The σ and its corresponding QY and its theoretical efficiency after DB calculation with NR impact under one sun illumination. Its maximum QY and its slopes are different than the case of $σ_1=4×10^{-3}$

|  | with $f_{NR}$ | | |
|---|---|---|---|
| $σ_1=2×10^{-3}$ | $f_{NR}$ | $E_g$ (eV) | Efficiency (%) |
| σ=0.01, $E_{th}$ = 2.03 $E_g$ | 0.200 | 1.20 | 34.7 |
| σ=0.1, $E_{th}$ = 2.4 $E_g$ | 0.020 | 1.30 | 29.1 |
| σ=0.2, $E_{th}$ = 2.8 $E_g$ | 0.010 | 1.37 | 27.6 |
| σ=0.3, $E_{th}$ = 3.1 $E_g$ | 0.0067 | 1.39 | 27.1 |
| σ=0.4, $E_{th}$ = 3.4 $E_g$ | 0.005 | 1.41 | 26.8 |
| σ=0.5, $E_{th}$ = 3.5 $E_g$ | 0.004 | 1.42 | 26.6 |

Table.2 optimum bandgaps and its maximum efficiencies with/without NR under one sun illumination

In order to verify NR model into DB, we test the relationships between Heaviside step function and Dirac delta function into the DB limit (see Eqns 11-16) under one sun illumination. And, we organize the simulation results in Fig.4 and 5 and Table 1 and 2 to compare the different case of QYs. Without NR, the efficiency of IQY case is similar to the ideal MEGSC. For NQY without NR, the performances of MEGSC becomes near the conventional single junction solar cell after $E_{th}=3E_g$ which its calculated efficiency is similar range of SQ limit. However, after comparing peak intensities by calculating $f_{NR}$, the overall efficiencies with NR is lower than SQ limit except for the case of σ=0.01. For instance, the theoretical maximum efficiency at σ=0.1 is 34.5% without NR and 30.8% with NR which $J_0$ should be lower than 25 times of $J_0$ to maintain the performances of MEGSC. And, its optimum bandgap with NR is 1.14 eV which it is similar to Silicon (see Fig.4 and Table 1). In other words, while regarding MEGSC with NR, the minimum requirement for generating MEG is that $E_{th}$ should be lower than $2.4E_g$ and its corresponding bandgap energy has to be lower than 1.14 eV. Furthermore, to compare the different slope of NQY, we additionally test by increased two times of $σ_1$. In this case, the overall performances except for σ=0.01 show negligible MEG effect that theoretical efficiencies are below SQ limit. Even if the calculated efficiency of NQY without NR at $E_{th}=3E_g$ shows slightly higher than SQ limit, its recombination term in DB limit considers only radiative recombination limit. Therefore, if there is a practical MEGSC with $E_{th}=3E_g$, its implied efficiency could be lower than 31.8% which its performance will be similar to conventional single junction solar cells.

## 4. Conclusions

Herein, we explained the mathematical approach of QY extraction based on the GDF to discuss the ideality of MEGSCs. Through the QY extraction analysis, first, we mathematically identified the limitations arising from material imperfections in MEGSCs owing to non-idealities. Based on the delta function analysis, we employed the GDF to determine the impact of EHP generation for each energy state of a MEGSC; furthermore, we analyzed the variation in the shape of the GDF distribution, and this distribution was rearranged. Based on the variation and rearrangement of the GDF distribution, we could theoretically determine the QY. When the peak intensities of the GDF between $E_g$ and $m·E_g$ (m=2,3,4…) are compared, the peak intensity difference can be employed to quantitatively analyze the occupancy of electrons at $mE_g$ and the non-idealities of MEGSCs. The large deviation from the ideal case was manifested via an increased $E_{th}$ and the linearly increasing slope of the QY plot. Furthermore, the non-idealities were explained based on the imperfections of the nanostructures. Notably, it is difficult to realize effective MEGSCs owing to the aforenoted imperfections, which lead to NR or non-idealities. Further, we apply these results into the DB limit of MEGSC with NR term by calculating the ratio between radiative and nonradiative recombination. With considering this ratio, the overall efficiencies are lower than SQ limit that its $E_{th}$ and optimum bandgaps are lower than $2.4E_g$ and 1.14 eV to maintain the characteristics of MEGSC when $σ_1 = 4 \times 10^{-3}$. Furthermore, to compare the different QY, we increase $σ_1$ by two times that its maximum QY is approximately 50% decreased and its theoretical performances also degraded. In other words, if the slope of NQY is decreased, its expected results become a single junction solar cell even if QY is slightly increased after $E_{th}$ (slope of NQY is 0.49).

**Author Contributions:** Dr. Jongwon Lee suggested the main idea, wrote, and reviewed this article.

**Funding:** N/A

**Acknowledgments:** N/A

11 of 11